\documentclass{article}
\usepackage{epsf}
\usepackage{graphicx}
\usepackage{amsmath}
\usepackage{amssymb}
\usepackage{epstopdf}
\usepackage{url}

\def\lsim{\mathrel{\raise.3ex\hbox{$<$\kern-.75em\lower1ex\hbox{$\sim$}}}}
\def\gsim{\mathrel{\raise.3ex\hbox{$>$\kern-.75em\lower1ex\hbox{$\sim$}}}}

\begin{document}

\pagenumbering{roman}
%\maketitle
%\tableofcontents
%\cleardoublepage
\pagenumbering{arabic}

%\part{Colliders searches}

\centerline{\Large \bfseries SUSY Tools for Dark Matter and at the Colliders}

\bigskip

\centerline{\large Fawzi Boudjema$^a$,
Joakim Edsj\"o$^b$, Paolo Gondolo$^c$}

\bigskip

{\em \noindent \small $^a$ LAPTH, Universit\'e de Savoie and CNRS, Chemin de Bellevue, F-74941 \\
Annecy-le-Vieux, France \\
$^b$ Oskar Klein Centre for Cosmoparticle Physics, Department of Physics, \\
AlbaNova, Stockholm University, SE-106 91 Stockholm, Sweden
\\ $^c$ Department of Physics, University of Utah,
   115 S 1400 E \# 201, Salt Lake City, UT 84112, USA}

\bigskip
\bigskip

\label{Chap:Boudjema}

\centerline{\large Chapter 16\footnote{This is a slightly modified version of our chapter in the published book.} of the book \emph{Particle Dark Matter},}
\centerline{Cambridge University Press, 2010, Editor: Gianfranco Bertone}
\bigskip
\centerline{Book webpage:}
\centerline{\url{http://cambridge.org/us/catalogue/catalogue.asp?isbn=9780521763684}}

\bigskip 
\bigskip

\begin{abstract}
With present and upcoming SUSY searches both directly, indirectly and at accelerators, the need for accurate calculations is large. We will here go through some
of the tools available both from a dark matter point of view and at accelerators. 
For natural reasons, we will focus on public tools, even though there are some rather
sophisticated private tools as well. 
\end{abstract}

\bigskip

The long awaited Large Hadron Collider (LHC) is expected to start taking data in 2009.
The LHC research program has
traditionally been centered around the discovery of the Higgs boson. However,
the standard model description of this particle calls for New
Physics. Until a few years ago, the epitome of this New Physics has
been supersymmetry, which when endowed with a discrete symmetry called R-parity furnishes a good dark matter candidate. Recently a few
alternatives have been put forward. Originally, they were confined to solving
the Higgs problem, but it has been discovered that, generically,
their most viable implementation (in accord with electroweak
precision data, proton decay, etc.) fares far better if a discrete
symmetry is embedded in the model. The discrete symmetry is behind the
existence of a possible dark matter candidate.

From another viewpoint, stressed in many parts of this book, the last few
years have witnessed spectacular advances in cosmology and
astrophysics confirming that ordinary matter is a minute part of
what constitutes the Universe at large. At the same time in which the
LHC will be gathering data, a host of non collider astrophysical
and cosmological observations with ever increasing accuracy will be carried out in search of
dark matter. For example, the
upcoming {\tt PLANCK} experiment will make cosmology enter the era
of precision measurements akin of what we witnessed with the LEP
experiments. 

The emergence of this new paradigm means it is of
utmost importance to analyse and combine data from these upcoming
observations with those at the LHC. This will also pave the way to
search strategies for the next Linear Collider, LC. This important
program is only possible if a cross-border particle-astroparticle
collaboration is set up having at its disposal common or
complementary tools to conduct global searches and analyses. Many
groups, from erstwhile diverse communities, are now developing,
improving, generalising, interfacing and exploiting such tools for
the prediction and analysis of dark matter signals from a
combination of terrestrial and non terrestrial observations,
paying particular attention to the astrophysical uncertainties.
Most of this work has been conducted in the context of
supersymmetry, but the latest numerical tools are not limited to it. 

In this chapter, we will go through some
of the tools available both from a dark
matter point of view and at accelerators. For natural reasons,
we will focus on public tools, even though there are some rather
sophisticated private tools as well. For supersymmetric dark matter
calculations, one of the first public tools available was {\tt
Neutdriver}~\cite{Jungman:1995df,Jungman:2000neut}. It was a
precursor in the field, but has by now been superceded by other
more sophisticated tools. There are currently three publicly available codes for calculations of dark matter densities and dark matter signals: {\tt
DarkSUSY}~\cite{Gondolo:2000ee,Gondolo:2002tz,Gondolo:2004sc}, {\tt
micrOMEGAs}~\cite{Belanger:2001fz,Belanger:2006is,Belanger:2004yn,Belanger:2008sj}
and {\tt IsaRED} (part of {\tt
ISASUSY/Isajet}~\cite{Paige:2003mg})

\section{Annihilation cross section and the relic density}

The general theory behind relic density calculations of dark
matter particles is given in Chapter 7 \cite{Chap:Gelmini}. Here
we will focus on supersymmetric dark matter (neutralinos) and various tools
available for calculating the relic density. There are currently
three publicly available codes for calculating the relic density
of neutralinos: {\tt
DarkSUSY}, {\tt
micrOMEGAs}
and {\tt IsaRED}\footnote{After this chapter was completed, a fourth code, {\tt SuperIso Relic} \cite{Arbey:2009gu} has appeared as well.}. All three of these codes are capable of reading (and sometimes writing)
SUSY Les Houches Accord (SLHA) files~\cite{Skands:2003cj,Allanach:2008qq} which allows for an easy interface between these codes and other tools to be described in Section~\ref{section-tools-collider}.

We will in the following
refer to these codes and how they calculate the relic density. We
will use the notation of Chapter 7 \cite{Chap:Gelmini} and only
write down the equations needed to facilitate our discussions
here.

%%%%%
\subsection{The Boltzmann equation}

In most supersymmetric models of interest for dark matter
phenomenology, the lightest neutralino, $\chi$ is the lightest
supersymmetric particle and our dark matter candidate. As such, we
want to calculate the relic density of neutralinos in the Universe
today as accurately as possible, which means that we need to solve
the Boltzmann equation.
\begin{equation}
  \frac{dn_\chi}{dt}=-3Hn_\chi-\langle \sigma_{\rm ann} v \rangle \left( n_\chi^2 - n_{\chi,\,\rm eq}^2 \right) \label{eq:boltzmann}
\end{equation}
where $n_\chi$ is the number density of neutralinos, $H$ is the
Hubble parameter, $\langle \sigma_{\rm ann} v \rangle$ is the
thermally averaged annihilation cross section and $n_{\chi,\,\rm
eq}$ is the equilibrium number density of neutralinos. This
equation needs to be solved over time (or temperature) properly
calculating the thermal average at each time step. When the
neutralinos no longer can follow the chemical equilibrium density
$n_{\chi,\,\rm eq}$, they are said to freeze-out. There are
several complications in solving Eq.~(\ref{eq:boltzmann}); for
example, we may have resonances and thresholds in our annihilation
cross section. The solution to this is to calculate the
annihilation cross section in general relativistic form, for
arbitrary relative velocities, $v$. Another complication is that
other supersymmetric particles of similar mass will be present
during freeze-out of the neutralinos. To solve this we need to
take into account the so-called coannihilations between all the
SUSY particles that are almost degenerate in mass with the
neutralino (in practice, it is often enough to consider
coannihilations between all SUSY particles up to about 50\%
heavier than the neutralino). Following the discussion in
Chapter~7~\cite{Chap:Gelmini}, we can solve for the total number
density of SUSY particles,
\begin{equation}
  n= \sum_{i=1}^N n_{i}
\end{equation}
instead of only the neutralino number density. It is also
advantageous to rephrase the Boltzmann equation in terms of the
abundance $Y=n/s$ and use $x=m_\chi/T$ as independent variable
instead of time or temperature. When coannihilations are included,
the Boltzmann equation 
(\ref{eq:boltzmann})
%~(\ref{eq:stdfi}) 
can then be written as
\begin{equation}
\frac{dY}{dx}=-\left(\frac{45}{\pi M_P^2}\right)^{-1/2}\frac{g_*^{1/2}m_\chi}{x^2}\left<\sigma_{\rm eff} v\right> (Y^2-Y_{eq}^2)\,.
\label{eq:Boltzco}
\end{equation}
where $\langle \sigma_{\rm eff} v \rangle$ is given by
\cite[Eq.~(7.18)]{Chap:Gelmini}
%Eq.~(\ref{eq:sigmavefffin2}).

%%%%%
\subsection{Solving the Boltzmann equation}
To solve the Boltzmann equation (\ref{eq:Boltzco}) we need to
calculate the thermally averaged annihilation cross section
$\langle \sigma_{\rm eff} v \rangle$ for each given time
(temperature). This is typically quite CPU-intensive, and we
therefore need to use some tricks. In {\tt DarkSUSY}
~\cite{Gondolo:2004sc}, the solution is speeded up by tabulating
$W_{\rm eff}$ in 
\cite[Eq.~(7.19)]{Chap:Gelmini}
%Eq.~(\ref{eq:weff})
, but using the momentum of
the $\chi$, $p_{\rm eff}$ as independent variable instead of $s$.
This tabulation takes extra care of thresholds and resonances
making sure that they are tabulated properly. This tabulated
$W_{\rm eff}$ is then used to calculate the thermal average
$\langle \sigma_{\rm eff} v \rangle$ for each time (temperature),
using 
\cite[Eq.~(7.18)]{Chap:Gelmini}
%Eq.~(\ref{eq:sigmavefffin2})
. The advantage with this method
is that $W_{\rm eff}$ does not depend on temperature and instead
the temperature dependence of $\langle \sigma_{\rm eff} v \rangle$
is completely taken care of by the other factors in
\cite[Eq.~(7.18)]{Chap:Gelmini}.
%Eq.~(\ref{eq:sigmavefffin2})
Numerically, one needs to take
special care of the modified Bessel functions $K_1$ and $K_2$
which both contain exponentials that need to be handled separately
to avoid numerical underflows. The Boltzmann Equation
(\ref{eq:Boltzco}) is then solved with a special implicit method with adaptive
stepsize control, which is needed because the equation is stiff and develops numerical instabilities unless an implicit method is used. The details of the DarkSUSY method are as follows (see {\tt DarkSUSY} manual~\cite{Gondolo:2004sc}). The derivative $dY/dx$ in Eq.~(\ref{eq:Boltzco}) is replaced with a finite difference $(Y_{n+1}\!-\!Y_n)/h$, where $Y_n=Y(x_n)$ and $x_{n+1}=x_n+h$. Then $Y_{n+1}$ is computed in two ways: first, the right hand side of Eq.~(\ref{eq:Boltzco}) is approximated with $-\tfrac{1}{2}[\lambda_{n}(Y_{n}^2-Y_{eq,n}^2)+\lambda_{n+1}(Y_{n+1}^2-Y_{eq,n+1}^2)]$, where $\lambda(x)=\left(45/\pi M_P^2\right)^{-1/2} (g_*^{1/2}m/x^2)\left<\sigma_{\rm eff} v\right>$, and an analytic solution is used for the resulting second-degree algebraic equation in $Y_{n+1}$; second, the right hand side of Eq.~(\ref{eq:Boltzco}) is approximated with $-\lambda_{n+1}(Y_{n+1}^2-Y_{eq,n+1}^2)$ and an analytic solution of the algebraic equation for $Y_{n+1}$ is used. The stepsize $h$ is reduced or increased to maintain the difference between the two approximate values of $Y_{n+1}$ within a specified error. Overall, the solution of the Boltzmann equation and the tabulation of $W_{\rm eff}$ solves for the relic density to within about 1\%. If needed, higher accurary can also be chosen as an option.

In {\tt micrOMEGAs}~\cite{Belanger:2006is}, $\langle \sigma_{\rm
ann} v \rangle$ at a given temperature $T$ is arrived at by
performing a direct integration and does not therefore rely on a
tabulation of the matrix elements squared. Two modes are provided
to perform the integration. In the  {\it accurate}  mode   the
program evaluates all integrals by means of an adaptive Simpson
integration routine. It automatically detects all singularities of
the integrands and checks the precision of the calculation
increasing the number of points until an accuracy of $10^{-3}$ is
reached. In the default mode ({\it fast} mode) the accuracy is not
checked but a set of points is sampled according to the behaviour
of the integrand: poles, thresholds and Boltzmann suppression at
high momentum. The first integral over scattering angles is
performed by means of a $5$-point Gauss formula. The accuracy of
this mode is generally about $1\%$. The user can also test the
precision of the approximation based on expanding the cross
section in terms of its $s$ and $p$-wave components.

In the Boltzmann Equation, we need to know $g_{\rm eff}(T)$ and
$h_{\rm eff}(T)$ that enter $g_*^{1/2}$ through 
\cite[Eq.~(7.9)]{Chap:Gelmini}.
%Eq.~(\ref{eq:gstar}).
In {\tt DarkSUSY}, the default is to use the
estimates in Ref.~\cite{Hindmarsh:2005ix}, but other options are
also available. Typically, different estimates of $g_{\rm eff}(T)$
and $h_{\rm eff}(T)$ translate into relic densities different by a
few percent.

{\tt IsaRED} on the other hand does not solve the
Boltzmann equation numerically, instead it finds the freeze-out
temperature (the temperature when the annihilation rate equals the
expansion rate of the Universe) and calculates the relic density
from that (including remnant annihilations at later times). For
the thermal averaged annihilation cross section, it uses the same
relativistic treatment as {\tt DarkSUSY} and {\tt micrOMEGAs}.

%%%%%
\subsection{Coannihilation criteria}

\begin{figure}
  \centerline{\includegraphics[width=0.75\textwidth]{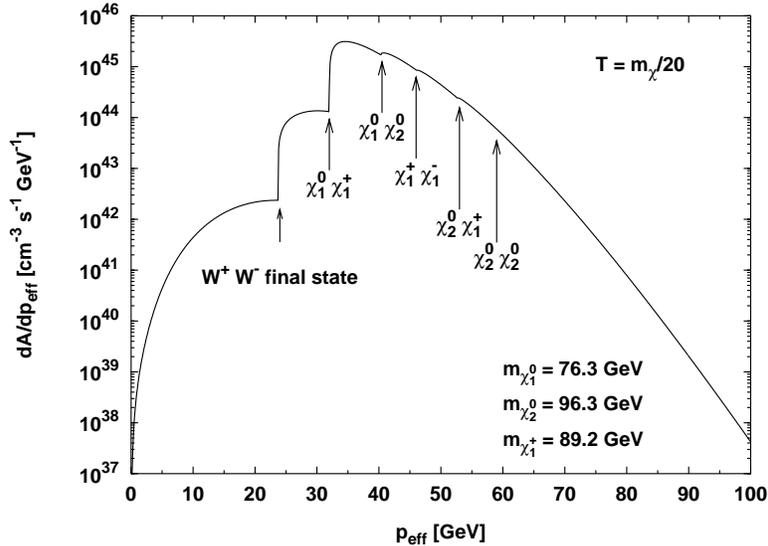}}
  \caption{Total differential annihilation rate per unit volume
    $dA/dp_{\rm eff}$ for the same model as in
    \cite[Fig.~7.3]{Chap:Gelmini},
%    Fig.~\protect\ref{fig:effrate}, 
    evaluated at a temperature
    $T=m_\chi/20$, typical of freeze-out. Notice the Boltzmann
    suppression at high $p_{\rm eff}$.}
  \label{fig:k1effrate}
\end{figure}

In principle, one should include all SUSY particle coannihilations
when calculating the relic density. However, the heavier they are,
the less abundant they will be and can thus be neglected. This is
important to speed up the calculation, as we would otherwise spend
most of our CPU cycles on calculating non-important coannihilation
cross sections. One can estimate which particles to include, by
investigating 
\cite[Eq.~(7.18)]{Chap:Gelmini}.
%Eq.~(\ref{eq:sigmavefffin2}). 
The modified Bessel
function $K_1$ contains an exponential, the so called Boltzmann
suppression, that will suppress all heavier particles. In
Fig.~\ref{fig:k1effrate}, we show $dA/dp_{\rm eff}$ for the same
model as in 
\cite[Fig.~7.3]{Chap:Gelmini}.
%Fig.~\ref{fig:effrate}.
$dA/dp_{\rm eff}$ is
essentially (apart from normalization) the integrand in the
numerator in 
\cite[Eq.~(7.18)]{Chap:Gelmini}.
%Eq.~(\ref{eq:sigmavefffin2}). 
Comparing the two
figures, we clearly see the Boltzmann suppression of larger
$p_{\rm eff}$, i.e.\ heavier coannihilating particles. We can
quantify this by comparing the Boltzmann suppression factor for
two coannihilating particles with masses $m_i$ and $m_j$ with the
corresponding factor for the LSP, the $\chi$. The suppression
factor for the coannihilating particles compared to the $\chi$ is
roughly (neglecting the $p_{\rm eff}^2$ in
\cite[Eq.~(7.18)]{Chap:Gelmini})
%Eq.~(\ref{eq:sigmavefffin2})) 
given by
\begin{equation} B =
\frac{K_1((m_i+m_j)/T)}{K_1(2m_\chi/T)} \simeq  e^{-x
\frac{m_i+m_j-2m_\chi}{m_\chi}} \quad  ; \quad x=\frac{m_\chi}{T}
\end{equation}
At freeze-out we typically have $x\simeq 20$, which
gives a suppression factor of $B\simeq 10^{-6}$ for coannihilation
particles about 50\% heavier than the LSP, $\chi$. In {\tt
DarkSUSY}, one can set the maximum mass fraction of
coannihilation particles, $f=m_i/m_\chi$ that will be included in
the calculation, whereas in {\tt micrOMEGAs} one sets the minimum
$B$ to allow instead. The defaults in {\tt DarkSUSY} ($f=1.5$) and
{\tt micrOMEGAs} ($B_{\rm min} = 10^{-6}$) are roughly equivalent.
One should remember, though, that the value to choose depends on
the particle physics model. For example, for chargino coannihilations,
the $\chi^+ \chi^-$ coannihilation cross section can be orders of
magnitudes larger than the $\chi \chi$ annihilation cross sections
and one should choose $f$ or $B_{\rm min}$ so that one does not
accidentally neglect coannihilations that are important. For the
MSSM, the default values of {\tt DarkSUSY} and {\tt micrOMEGAs}
are typically sufficient for all interesting cases. {\tt IsaRED}
instead includes a preset collection of particles that are of
relevance for the mSUGRA setup.

%%%%%
\subsection{Annihilation cross section}

At the heart of the relic density calculation are the annihilation
and coannihilation cross sections. In the MSSM there are over 2800 sub-processes (not counting charged-conjugate final
states) that can in principle contribute in the relic density calculation. It appears at first sight to be a daunting task to
provide such a general code.

In {\tt DarkSUSY}, all annihilation and coannihilation cross
sections for the MSSM\footnote{Gluino coannihilations are currently not
included.} are calculated at tree level by hand with
the help of symbolic programs like {\tt Reduce}, {\tt Form} or
{\tt Mathematica}. The calculations are performed with general
expressions for the vertices in the Feynman rules and the
results are converted to {\tt Fortran} code. The vertices are then
calculated numerically for any given MSSM model. The analytically
calculated cross sections are differential in the angle of the
outgoing particles, and the integration over the outgoing angle is performed numerically.

In {\tt micrOMEGAs}, on the other hand, any annihilation and
coannihilation cross sections are calculated automatically and
generated on the fly. This is possible thanks to an
interface to {\tt CalcHEP}~\cite{Pukhov:2004ca}, which is an
automatic matrix element/cross sections generator. This automation
is carried one step further in that {\tt CalcHEP} itself reads its
MSSM model file (Feynman rules) from {\tt
LanHEP}~\cite{Semenov:1998eb}, which outputs the complete set of
Feynman rules from a simple coding of the Lagrangian, see
section~\ref{section-tools-collider}. In the first call to {\tt
micrOMEGAs} only those subprocesses needed for the given
set of the MSSM parameters are generated. The corresponding
``shared"  library  is stored on the user disk space and is
accessible for all subsequent calls, thus each process is
generated and compiled only once. This library is then filled with
more and more processes whenever the user needs new processes for
different MSSM scenarios. This avoids having to distribute a huge
code with all the possible 2800 processes.

Both methods have advantages and disadvantages. In the {\tt
DarkSUSY} setup, no recalculation of the (analytical)
annihilation cross sections is needed, which can speed things up.
Also, the analytically calculated annihilation cross sections can
be optimized to be faster. On the other hand, the {\tt micrOMEGAs}
setup makes it easier to adapt the code to non-MSSM cases. In
both codes though, the actual Boltzmann equation solver is very
general and works for any kind of WIMP dark matter, not only SUSY
dark matter. In {\tt IsaRED}, {\tt CompHEP}~\cite{Boos:2004kh} is
used to calculate the annihilation and coannihilation cross
sections for a subset of SUSY particles of relevance mostly for
mSUGRA (the two lightest neutralinos, the lightest chargino, the
left-handed eigenstates of sleptons and squarks, and gluinos).  The expressions for the annihilation cross sections in {\tt IsaRED} are not calculated on the fly, but are instead precalculated and included with the code.

%%%%%
\section{Direct detection}

Detailed expressions for detection rates in direct detection experiments are presented in Chapter~17~\cite{Chap:Cerdeno}. Here we focus on characteristics of the elastic scattering cross sections and event rates as they are implemented in numerical tools.

Direct detection rates depend on the differential elastic WIMP-nucleus cross section $d\sigma_{WN}/dE_{R}$, where $E_{R}$ is the energy of the recoiling nucleus:
\begin{equation}
\frac{d\sigma_{WN}}{dE_{R}} = \frac{m_N}{2 \mu_N^2 v^2} \Bigl( \sigma_0^{SI} F^2_{SI}(E_R) + \sigma_0^{SD} F^2_{SD}(E_R)\Bigr) .
\end{equation}
Here $m_N$ is the nucleus mass, $\mu_N=m_\chi m_N/(m_\chi+m_N)$ is the WIMP-nucleus reduced mass ($m_\chi$ being the WIMP mass), ${\bf v}$ is the WIMP-nucleus relative velocity before the collision, $\sigma_0^{SI,SD}$ are the spin-independent and spin-dependent cross sections at zero momentum transfer, and $F^2_{SI,SD}(E_R)$ are the squares of the corresponding form factors (also called structure functions). In terms of these quantities, the directional and non-directional direct detection rates read
\begin{eqnarray}
\frac{dR}{dE_Rd\Omega_q} &=&  \frac{\rho_0}{8 \pi \mu_N^2 m_\chi} \,  \sigma_{WN}(E_R) \,\hat{f}(v_{\rm min},\Omega_q),
\label{eq:dRdEdO}\qquad
\\
 \frac{dR}{dE_R} &=&
 \frac{\rho_0}{2 \mu_N^2 m_\chi} \,  \sigma_{WN}(E_R)\, \eta(v_{\rm min}) ,
\label{eq:dRdE}\qquad
\end{eqnarray}
where
\begin{eqnarray}
\sigma_{WN}(E_R) & = & \sigma_0^{SI} F^2_{SI}(E_R) + \sigma_0^{SD} F^2_{SD}(E_R), \\
\eta(v_{\rm min}) & = & \int_{v>v_{\rm min}} \frac{f({\bf v})}{v} \, d^3 v ,
\end{eqnarray}
$v_{\rm min}=[(m_N E_R)/(2\mu_N^2)]^{1/2}$ is the minimum WIMP speed which can cause a recoil of energy $E_{R}$, $\Omega_q$ is the direction of the nucleus recoil momentum of magnitude $q=\sqrt{2 m_N E_R}$, and $\hat{f}(v_{\rm min},\Omega_q)$ is the Radon transform of the velocity distribution function $f({\bf v})$.
%Eq.~(\ref{eq:dRdE}) generalizes 
%\cite[Eq.~(17.1)]{Chap:Cerdeno}
%%Eq.~(\ref{diff_rate})
%of Chapter~17~\cite{Chap:Cerdeno} to anisotropic WIMP velocity distributions $f({\bf v})$, %and Eq.~(\ref{eq:dRdEdO}) generalizes 
%\cite[Eq.~(17.28)]{Chap:Cerdeno}
%%Eq.~(\ref{directional_diff_rate})
%to also include spin-dependent cross sections.

An important property of Eqs.~(\ref{eq:dRdEdO}) and~(\ref{eq:dRdE}) is the factorization of the particle physics properties, $\sigma_{WN}(E_R)$, and the astrophysics properties, $\rho_0 \hat{f}(v_{\rm min},\Omega_q)$ and $\rho_0 \eta(v_{\rm min})$.

Dark matter codes such as {\tt
DarkSUSY}, {\tt
IsaRED/RES} and
{\tt micrOMEGAs} 
 compute the particle physics and the astrophysics factors to various levels of precision and offer a number of choices for the form factors and the velocity distribution (more and more as they are upgraded). All codes provide the zero-momentum transfer cross sections (although some are still limited to the axial and scalar couplings of supersymmetric neutralinos), and most of the codes provide routines for direct detection rates off composite targets besides single nuclei.

For the spin-independent part, dark matter codes use the factorized form $\sigma_{WN}^{SI}(E_R) = \sigma_0^{SI} F^2_{SI}(E_R) $ with, in the notation of Chapter~17~\cite{Chap:Cerdeno},
\begin{equation}
\sigma_0^{SI} = \frac{4\mu_n^2}{\pi} \left[ \Bigl( Z f^p + (A-Z) f^n\Bigr)^2 + \frac{B_N^2}{256} \right] .
\end{equation}
Various expressions for the spin-independent form factor $F^2_{SI}(E_R)$ are typically available. For example, {\tt DarkSUSY} 5 automatically selects the best available form factor among Sums-of-Gaussians, Fourier-Bessel, and Helm parametrizations (see~\cite{Duda:2006uk} for a comparison of these approximations).

The spin-dependent part is often not factorized, so as to use the same functions provided by detailed simulations at zero and non-zero momentum transfer. With the by-now-standard normalization of the spin structure functions in~\cite{Engel:1991wq}, one has
\begin{eqnarray}
\sigma_{0}^{SD} F^2_{SD}(E_R) & = & \frac{32\mu_{N}^{2}G_{F}^{2}}{2J+1}\,\left[a_{0}^{2}S_{00}(E_R) + a_{1}^{2}S_{11}(E_R) + a_{0}a_{1}S_{01}(E_R)\right] \qquad{~} \\
& = & \frac{32\mu_{N}^{2}G_{F}^{2}}{2J+1}\,\left[a_{p}^{2}S_{pp}(E_R) + a_{n}^{2}S_{nn}(E_R) + a_{p}a_{n}S_{pn}(E_R)\right] ,\qquad{~}
\label{eq:sigmanSD}
\end{eqnarray}
where $a_0=a_p+a_n$, $a_1=a_p-a_n$,
\begin{eqnarray}
S_{{pp}}(E_R) & = & S_{00}(E_R)+S_{11}(E_R)+S_{01}(E_R)\,,
\\
 S_{{nn}}(E_R) & = & S_{00}(E_R)+S_{11}(E_R)-S_{01}(E_R)\,,
\\
S_{pn}(E_R) & = & 2\left[S_{00}(E_R)-S_{11}(E_R)\right]\,.
\label{Ss}
\end{eqnarray}
When the nuclear spin is approximated by the spin of the odd nucleon only, one finds~\cite{Alenazi:2007sy}
\begin{equation}
S_{{pp}}(E_R) = \frac{\lambda_N^2 J(J+1)(2J+1)}{\pi},\quad S_{nn}(E_R) = 0, \quad S_{{pn}}(E_R) = 0,
\label{Spp}
\end{equation}
for a proton-odd nucleus, and
\begin{equation}
S_{{pp}}(E_R) = 0, \quad S_{{nn}}(E_R) = \frac{\lambda_N^2 J(J+1)(2J+1)}{\pi}, \quad S_{pn}(E_R) = 0,
\end{equation}
for a neutron-odd nucleus. Here $\lambda_{N}$ is conventionally defined through the relation $\langle N|{\bf S}|N\rangle = \lambda_{N} \langle N|{\bf J}|N\rangle$, where $|N\rangle$ is the nuclear state, ${\bf S}$ is the nuclear spin, ${\bf J}$ is the nuclear total angular momentum. Tables of $\lambda_{N}^{2}J(J+1)$ values for several nuclei can be found in~\cite{Ellis:1991ef} and~\cite{Smith:1988kw}.

The quantities $f^p$, $f^n$, $B_N$, $a_0$, $a_1$ are sums of products of the WIMP-quark and WIMP-gluon coupling constants $\alpha_{q,G}^{S,V,A,P,T}$ (for scalar, vector, axial, pseudoscalar, and tensor currents) and of the contributions $f_{TG}$, $f_{Tq}$ and $\Delta_q$ of the gluons and each quark flavor to the mass and spin of protons and neutrons. Values for the nucleonic matrix elements of gluons and quarks, in practice values for $f_{TG}$, $f_{Tq}$ and $\Delta_q$, are either hardcoded or settable by the user. Values for the effective coupling constants $\alpha_{q}^{S,V,A,P,T}$ are either precomputed analytically ({\tt DarkSUSY}) or computed numerically on the fly ({\tt micrOMEGAs}). For example, the effective lagrangian at the zero momentum transfer for the interaction of a fermionic WIMP $\chi$ with quarks $q$ reads
\begin{eqnarray}
\label{fermion_si_lgrgn}
  {\cal L}_{F} &=&  \alpha_q^S \; \bar \chi \chi \; \bar q q +
\alpha_q^V \; \bar \chi \gamma_\mu \chi  \; \bar q \gamma^\mu q\nonumber\\
 &+& \alpha_q^P \; \bar \chi \gamma_5\gamma_{\mu} \chi   \;
 \bar q \gamma_5\gamma^{\mu}q
 +\tfrac{1}{2} \alpha_q^T \; \bar \chi \sigma_{\mu\nu} \chi \;  \bar q
 \sigma^{\mu\nu}q .
\end{eqnarray}
In the case of a Majorana WIMP, like the neutralino in the MSSM,
only operators even under $\chi \leftrightarrow \bar \chi$ are possible (i.e.\ $\alpha_q^V=\alpha_q^T=0$). In {\tt micrOMEGAs}, the numerical values of the coefficients $\alpha_q$ are obtained combining appropriate matrix elements for $\chi q \to \chi q$ and $\bar\chi q \to \bar \chi q$ scattering at zero momentum transfer. For example,
\begin{eqnarray}
\label{MEplus}
 \alpha_q^S + \alpha_q^V = - i\,
\frac{\langle q(p_1),\chi(p_2)| \hat{S} {\mathcal{O}_S} |q(p_1),\chi(p_2) \rangle }{ \langle q(p_1),\chi(p_2)|
{\mathcal{O}_S} {\mathcal{O}_S}
|q(p_1),\chi(p_2)\rangle}\nonumber
\\
\label{MEminus}
 \alpha_q^S - \alpha_q^V = - i\,
\frac{\langle \bar{q}(p_1),\chi(p_2)| \hat{S} {\mathcal{O}_S} |\bar{q}(p_1),\chi(p_2) \rangle }{ \langle
\bar{q}(p_1),\chi(p_2)| {\mathcal{O}_S} {\mathcal{O}_S} |\bar{q}(p_1),\chi(p_2)\rangle}
\end{eqnarray}
where $\mathcal{O}_S=\bar \chi \chi \, \bar q q$, the $S$-matrix $\hat S=1-i {\cal L}$ is obtained from the
complete Lagrangian at the quark level, and the scattering matrix elements on the right hand sides are computed with {\tt CalcHEP}. More general cases, including a generic local WIMP-quark operators and WIMPs with spin-$0$ and
spin-$1$, are presented in~\cite{Belanger:2008sj}.

Loop contributions are essential in the treatment of the WIMP-quark and especially WIMP-gluon coupling constants. For example, for neutral WIMPs like the supersymmetric neutralino, there is no neutralino-gluon coupling at the tree-level and the gluon contribution to $\alpha^S$ arises at the one-loop level. Complete analytic one-loop calculations for neutralino-quark and neutralino-gluon couplings were performed in~\cite{Drees:1992rr,Drees:1993bu}; these formulas are incorporated in {\tt DarkSUSY}. Automatic numerical calculations of all $\alpha_q^{S,V,A,P,T}$ at one-loop from user-specified generic lagrangians (with approximate treatment of some of the loop corrections, see~\cite{Belanger:2008sj}) are currently available in {\tt micrOMEGAs}.

%%%%%

\section{Indirect detection}

Indirect detection methods are many and varied. Here we focus on the following traditional methods: neutrinos from the Sun and the Earth, and gamma-rays, neutrinos, and charged cosmic rays (positrons, antiprotons and antideuterons) from annihilations in the galactic halo. There are also other indirect signals, like synchrotron emission, signals from cosmological halos (giving a diffuse flux), and indirect consequences of the presence of dark matter in stars, Chapter~29~\cite{Chap:Bertone}, but we will not focus on them here. Most of the theory needed for this discussion is found in Chapters~24~\cite{Chap:Bergstrom}, 25 \cite{Chap:Halzen} and 26 \cite{Chap:Salati}; we use the notation in those chapters and elaborate on the formulae given there when needed.

The main public tools available to calculate indirect rates are {\tt
DarkSUSY}~\cite{Gondolo:2000ee,Gondolo:2002tz,Gondolo:2004sc} and {\tt micrOMEGAs}~\cite{Belanger:2001fz,Belanger:2006is,Belanger:2004yn,Belanger:2008sj}. In addition, there are also approximate simple formulae and parameterizations available that can be used, but we will here focus on the numerical codes.

%%%%%
\subsection{Neutrinos from the Sun/Earth}

To calculate the neutrinos from the Sun/Earth, we need to calculate the capture rate of neutralinos in the Sun/Earth, we then need to solve the evolution equation for capture and annihilation in the Sun/Earth, let the neutralinos annihilate in the centre of the Sun/Earth to produce neutrinos and finally let the neutrinos propagate  to the neutrino detector at Earth (taking interactions and oscillations into account).

In Chapter~25~\cite{Chap:Halzen}, approximate formulae are given for the capture rate in the Sun. These formulae are good for quick calculations, but they include several approximations; with numerical codes, we can actually do better. {\tt DarkSUSY} is currently the only public code that includes neutrino fluxes from the Sun/Earth and uses the full expressions in Ref.~\cite{Gould:1987ir}, where the capture is integrated over the full Sun/Earth including capture on the 16 main elements for the Sun (and 11 for the Earth). In {\tt DarkSUSY}, an arbitrary velocity distribution can be used if desired in place of the commonly assumed Maxwell-Boltzmann distribution. For example, the Earth does not capture WIMPs directly from the galactic halo, instead it captures from a distribution that has diffused into the solar system by gravitation interactions \cite{1991ApJ...368..610G}. {\tt DarkSUSY} uses a velocity distribution at the Earth based on numerical simulations that take this diffusion into account \cite{Lundberg:2004dn}. There are also indications from more recent numerical simulations of WIMP diffusion in the solar system~\cite{Peter:2008sy} that heavier WIMPs will have a reduced capture rate in the Sun due to gravitational effects due to Jupiter. The {\tt DarkSUSY} user can optionally include these effects.

After capture, the evolution equation for the number density accumulated in the Sun/Earth is solved to give the annihilation rate today. Once the WIMPs have accumulated in the centre of the Sun/Earth, they annihilate and eventually produce neutrinos. In {\tt DarkSUSY}, the annihilation and propagation of neutrinos is handled by a separate code, {\tt WimpSim}~\cite{Blennow:2007tw,Edsjo:2008WimpSim}. {\tt WimpSim} takes care of annihilations to standard model particles in the central regions of the Sun/Earth with the help of {\tt Pythia}~\cite{Sjostrand:2006za}. Energy losses and stopping of particles in the dense environments at the center of Sun/Earth are also included. All flavours of neutrinos (and antineutrinos) are then propagated out of the Sun/Earth, taking oscillations and interactions (the latter only relevant for the Sun) into account. This is done in a full three-neutrino-flavour setup~\cite{Blennow:2007tw}. Once at the detector, the neutrinos are let to interact and produce charged leptons and hadronic showers. {\tt WimpSim} has been run for a range of annihilation channels and masses from 10 GeV to 10 TeV, and the results have been summarized as yield tables that are read and interpolated by {\tt DarkSUSY}. These results agree very well with a similar analysis in Ref.~\cite{Cirelli:2005gh}, where parameterizations and downloadable data files are also given. For annihilation channels that are particle physics model dependent (like annihilation to Higgs bosons), the Higgs bosons are let to decay in flight in {\tt DarkSUSY} and the resulting fluxes are calculated from their decay products (properly Lorentz boosted).

The routines in {\tt DarkSUSY} are also general enough to be easily adapted to other particle dark matter candidates, like Kaluza Klein dark matter.

%%%%%
\subsection{Charged cosmic rays}

The theory behind propagation of charged cosmic rays in the galaxy is presented in Chapter~26~\cite{Chap:Salati}. We will use the notation in that chapter. In principle, what we have to do is to solve the master equation 
\cite[Eq.~(26.4)]{Chap:Salati}
%(\ref{master_equation}) 
with appropriate diffusion coefficients, energy loss terms, source terms and boundary conditions. Currently, {\tt micrOMEGAs} includes the source spectra for arbitrary SUSY models, but does not include the spectra after propagation. This will be included in future versions though, using the results in Ref.~\cite{Brun:2006cj}. In {\tt DarkSUSY}, both the source spectra and the spectra after propagation in various propagation models are included. {\tt DarkSUSY} implements axisymmetric propagation models and spherically symmetric (or at least axisymmetric) halo models. In {\tt DarkSUSY}, diffusion is assumed to take place only in space (i.e.\ the term $K_{EE}$ in 
\cite[Eq.~(26.4)]{Chap:Salati})
%Eq.~(\ref{master_equation})
is assumed to be negligible). However, it also offers a full interface and integration with the leading cosmic ray propagation code {\tt GALPROP}~\cite{Strong:2001fu} where more sophisticated propagation models can be used.
For the halo density, several preset profiles are available, the default being an NFW profile~\cite{Navarro:1995iw}. However, the user can supply her/his own halo profile; if it is given in the form of 
\cite[Eq.~(26.15)]{Chap:Salati}
%Eq.~(\ref{eq:DM_profile_a}), 
it is particularly simple to do so. For solar modulation, {\tt DarkSUSY} offers a standard spherical force-field approximation as explained in Chapter~26~\cite{Chap:Salati}.

For the source spectra (i.e.\ the spectra before propagation), {\tt DarkSUSY} uses a similar setup as for the neutrino fluxes from the Sun/Earth given above. A large set of annihilation channels are simulated with {\tt Pythia}~\cite{Sjostrand:2006za} in vacuum, for a range of masses, and the yields of antiprotons, positrons, gamma-rays and neutrinos are stored as data tables. These tables are then read and interpolated by {\tt DarkSUSY} at run-time. Higgs boson decays are included stepping down the decay chain. {\tt micrOMEGAs} currently uses the same data files as {\tt DarkSUSY}, but both codes are planning on using new updated simulations in future releases.

For more details about the {\tt DarkSUSY} implementation of antiproton fluxes, see Ref.~\cite{Bergstrom:1999jc}; for the positron fluxes, see Ref.~\cite{Baltz:1998xv}. The antideuteron fluxes are calculated from the antiproton fluxes with the method given in Ref.~\cite{Donato:1999gy}.

%%%%%
\subsection{Gamma-rays and neutrinos}

Gamma-ray and neutrino spectra from annihilation in the galactic halo can be calculated with both {\tt DarkSUSY} and {\tt micrOMEGAs}. As mentioned above, the {\tt DarkSUSY} spectra are based upon {\tt Pythia} simulations, which are then read in and interpolated by {\tt DarkSUSY}. Currently, {\tt micrOMEGAs} uses the same tables as {\tt DarkSUSY}. {\tt DarkSUSY} also includes internal bremsstrahlung photons~\cite{Bringmann:2007nk} that can be very important in some parts of the parameter space (e.g.\ in the stau coannihilation region). {\tt DarkSUSY} also includes the monochromatic gamma-ray lines from annihilation to $\gamma \gamma$~\cite{Bergstrom:1997fh} and $Z \gamma$~\cite{Ullio:1997ke} that occur at loop-level. Currently, the {\tt DarkSUSY} (and consequently then also the {\tt micrOMEGAs}) neutrino spectra from halo annihilations do not include neutrino oscillations, but this will be addressed in future versions.

Gamma-rays and neutrinos are not affected by propagation, and hence the flux can be written as
\begin{eqnarray}
\frac{d^2\Phi_{\gamma/\nu}(\psi)}{dE d\Omega}
& \simeq  & 9.395 \cdot 10^{-12}\left( \frac{dN_{\gamma/\nu}}{dE}\right) \left(\frac{\sigma v}
{10^{-29}\ {\rm cm}^3 {\rm s}^{-1}}\right)\left( \frac{10\,\rm{GeV}}
{m_\chi}\right)^2 \nonumber \\
& & \times
J\left(\psi\right)\;\rm{cm}^{-2}\;\rm{s}^{-1}\;\rm{sr}^{-1}\;,
\end{eqnarray}
where we have defined the dimensionless function
\begin{equation}
J\left(\psi\right) = \frac{1} {8.5\, \rm{kpc}}
\cdot \left(\frac{1}{0.3\,{\rm GeV}/{\rm cm}^3}\right)^2
\int_{\rm line\;of\;sight}\rho_{\chi}^2(l)\; d\,l(\psi)\;
\label{eq:jpsi}
\end{equation}
with $\psi$ being the angle from the galactic centre direction to the direction of observation, and $dN_{\gamma/\nu}/dE$ being the spectrum of gamma-rays or neutrinos. The line of sight integral, Eq.~(\ref{eq:jpsi}), can be calculated for any spherically symmetric profile in {\tt DarkSUSY} (whereas {\tt micrOMEGAs} currently implements an isothermal profile only).

%%%%%
\section{Exploring the parameter space}

One problem that arises when exploring a specific supersymmetric model setup (e.g.\ mSUGRA or a low-energy MSSM model) is how to scan the parameter space. The dimensions of this parameter space, i.e.\ the number of free independent parameters, can be large. For example, the general MSSM model has 124 parameters (MSSM-124), 18 of which define the Standard Model (SM). If one assumes CP-conservation, the number of parameters reduces to 63 (MSSM-63), which is still a large number. Typically, one reduces the number of parameters still further with inspired theoretical insights (see Chapter~8~\cite{Chap:Ellis}). In Minimal Supergravity, for example, unification of coupling constants, of gaugino masses, and of scalar masses leaves only 23 parameters, i.e.\ SM+5, one of which is just a positive or negative sign. An intermediate model often used in neutralino dark matter studies (MSSM-25) has 25 parameters, i.e.\ SM+7.

One typically wants to find parameter values that are theoretically consistent, have a preferred relic density and are not already excluded by other searches (e.g.\ rare decays or other accelerator searches). The brute force method would be to scan over the parameter space with some kind of grid scan. One soon realizes that one can typically get a better efficiency in the scans (i.e.\ more points that pass the cuts, or a better sampling of different interesting regions in the parameter space) by scanning in the logarithm of the mass parameters instead of the mass parameters directly. For higher-dimensional parameter spaces, it is often also more advantageous to scan randomly instead of on a fixed grid.

\begin{figure}
  \centerline{\includegraphics[width=0.6\textwidth]{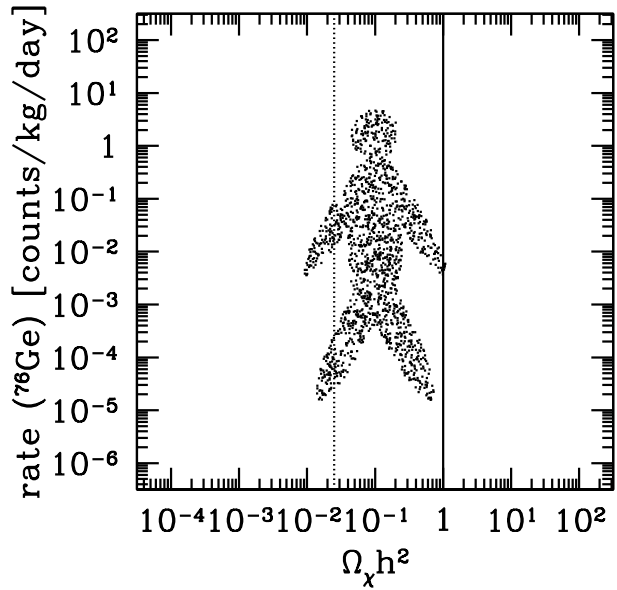}}
  \caption{A humorous scatter plot showing correlations that may arise with appropriate choices of priors while scanning parameter space: in this case, the priors reflect the anthropic principle.}
  \label{fig:littleman}
\end{figure}

However, none of these methods are very effective in finding regions of parameter space that pass all the cuts. The relic density cut alone discards many models because of the high precision with which we know the relic density of dark matter today. Hence, more sophisticated methods have evolved that are efficient in generating points inside the interesting regions. Most of these use a Markov Chain Monte Carlo (MCMC)~\cite{Gilks:1996book,Dunkley:2004sv,Baltz:2004aw,Allanach:2005kz,deAustri:2006pe,Baltz:2006fm} to generate points according to a goal distribution specified by the researcher. The goal distribution can be a function without direct physical meaning (e.g.\ a Gaussian distribution that peaks in the desired region) or could have a statistical meaning (e.g.\ a likelihood function or a prior distribution in a Bayesian analysis). A recent public code to perform these tasks, and that is linked to {\tt DarkSUSY}, {\tt micrOMEGAs} and other codes, is {\tt SuperBayeS}~\cite{deAustri:2006pe}. These advanced methods can be very effective in finding the interesting regions of the parameters space. However, when interpreting the distribution of points these methods produce, one should be very careful. The definition of ``interesting'' is different for different investigators, and the way points are generated always involves a prior in parameter space (even grid methods can be said to have a prior, namely a series of Dirac delta functions at each grid point). One could go to the extreme of producing any kind of results by choosing appropriate priors. Fig.~\ref{fig:littleman}, for example, shows the correlation between the direct detection rate and the relic density jokingly obtained with priors that reflect the anthropic principle.
When parameter space scans with priors are used to compute statistical inferences on data, e.g.\ likelihood contours~\cite{Trotta:2008bp}, one should keep in mind their rather severe dependence on the priors, especially the priors on very poorly known parameters for which there are little or no experimental data (supersymmetric masses, e.g.). This dependence arises from the use of Bayes theorem, which gives the probability of the model parameter given the data (the likelihood) in terms of the probability of the data given the model parameters (the assumed distribution of experimental and theoretical errors) as
\begin{equation}
{\rm Prob(model|data)} = {\rm Prob(data|model)} \; \frac{{\rm Prob(model)}}{{\rm Prob(data)}} .
\end{equation}
While the probability of the data Prob(data) is just a normalization constant, the probability of the model parameters Prob(model) is the prior representing the degree of belief or the relative preference the researcher has in specific values of the model parameters. When real experimental data on particle dark matter models are in, the dependence on the priors is expected to become less severe.

%%%%%
\section{Interface with collider and precision measurements codes}
\label{section-tools-collider}

\begin{figure*}%[thbp]
\begin{center}
\includegraphics[width=0.7\textwidth,height=\textwidth,angle=-90]{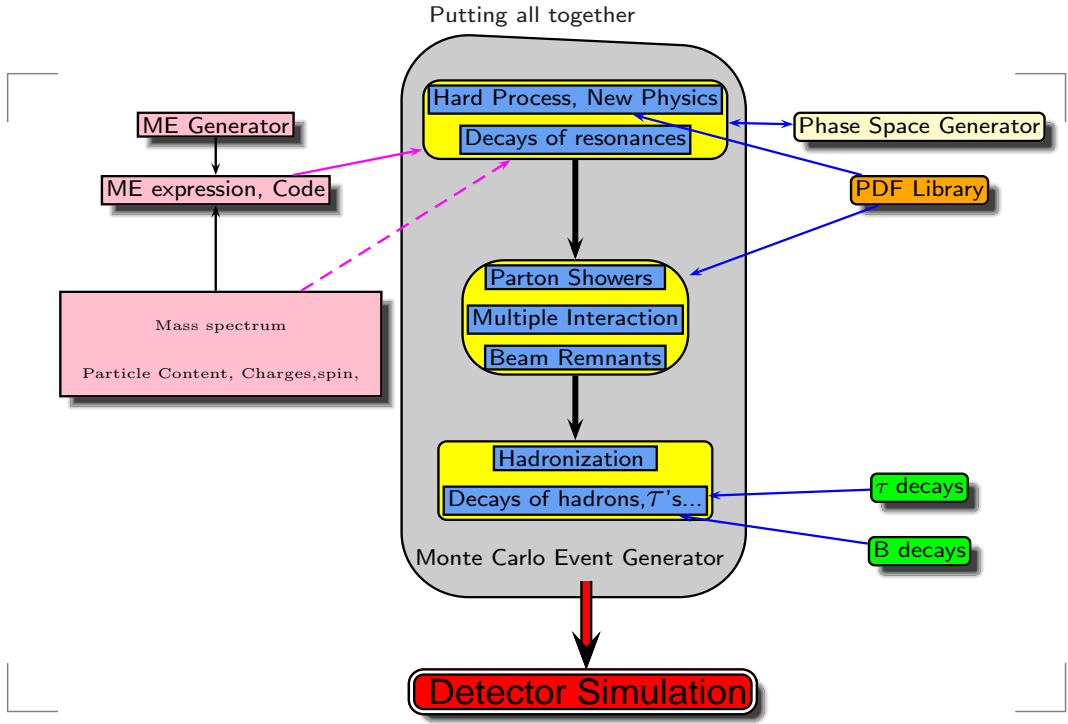}
\caption{\label{mc-generators}{Modules that are involved in
the construction of a Monte-Carlo event generator for particle
collider physics.}}
\end{center}
\end{figure*}

Until a few years ago, one used constraints on the
inferred amount of dark matter to delimit the parameter space of
supersymmetry in order to narrow  searches of supersymmetry at the
colliders. Now with the improvement on the precision on the
cosmological parameters one asks whether the LHC and LC can match
the precision of the upcoming cosmology and astrophysics
experiments in, for example, reconstructing the relic density once
supersymmetry is identified~\cite{Allanach:2004xn,Baltz:2006fm}.
This may even bring a bonus in that one can test some cosmological
and astrophysical assumptions, like indirectly ``probing" the
history of the early universe. For this programme to be feasible
one needs to control the particle physics component with as much
accuracy as possible. To be able to conduct a cohesive and
self-consistent precision test of the origin of DM from the
particle physics point of view one needs to calculate not only
those dark matter cross sections but also the observables at the
colliders that are predicted for the same dark matter model.
Ideally, therefore, one would like a common tool that performs this
task. In many instances this also requires that one goes beyond
calculations at tree-level. This is especially true in the case of
supersymmetry where it is known that radiative corrections can be
large. Some progress in this direction has also been made recently
within the {\tt SloopS}
Collaboration~\cite{Boudjema:2005hb,Boudjema:2006kn,Baro:2007em,Baro:2008bg,Freitas:2004kh}.

The dark matter codes for supersymmetry such as {\tt
DarkSUSY}, {\tt
IsaRED/RES}  and
{\tt micrOMEGAs}
include some higher order effects. Moreover, because of the
complexity of the MSSM which has a large number of parameters and
a large array of predictions, these codes also rely on other more
specific codes that predict various other observables in
supersymmetry. This concerns for example  codes for the
calculation of the spectrum based on the renormalisation group
equations (RGE) that predict the low energy physical masses from
an input at the unification scale in some constrained model of
supersymmetry breaking. Spectrum calculators include {\tt Spheno}~\cite{Porod:2003um}, {\tt Softsusy}~\cite{Allanach:2001kg}, {\tt
suspect}~\cite{Djouadi:2002ze} and {\tt ISASUGRA} (part of {\tt ISAJET})~\cite{Paige:2003mg}. These codes themselves may borrow
from more specialised codes like those for the calculation of the
Higgs masses, such as {\tt Feynhiggs}~\cite{Heinemeyer:1998yj}.
The codes for the mass spectra may also feed in stand-alone
codes for the calculation of precision measurements, like the
calculation of $(g-2)_{\mu}$ and $B$ observables. Example of such
codes or ``flavour calculators'' are ({\tt
SUSYbsg}~\cite{Degrassi:2007kj}, {\tt superiso}~\cite{Mahmoudi:2007vz}).  

\begin{figure*}%[htbp]
\begin{center}
\includegraphics[width=0.7\textwidth,height=\textwidth,angle=-90]{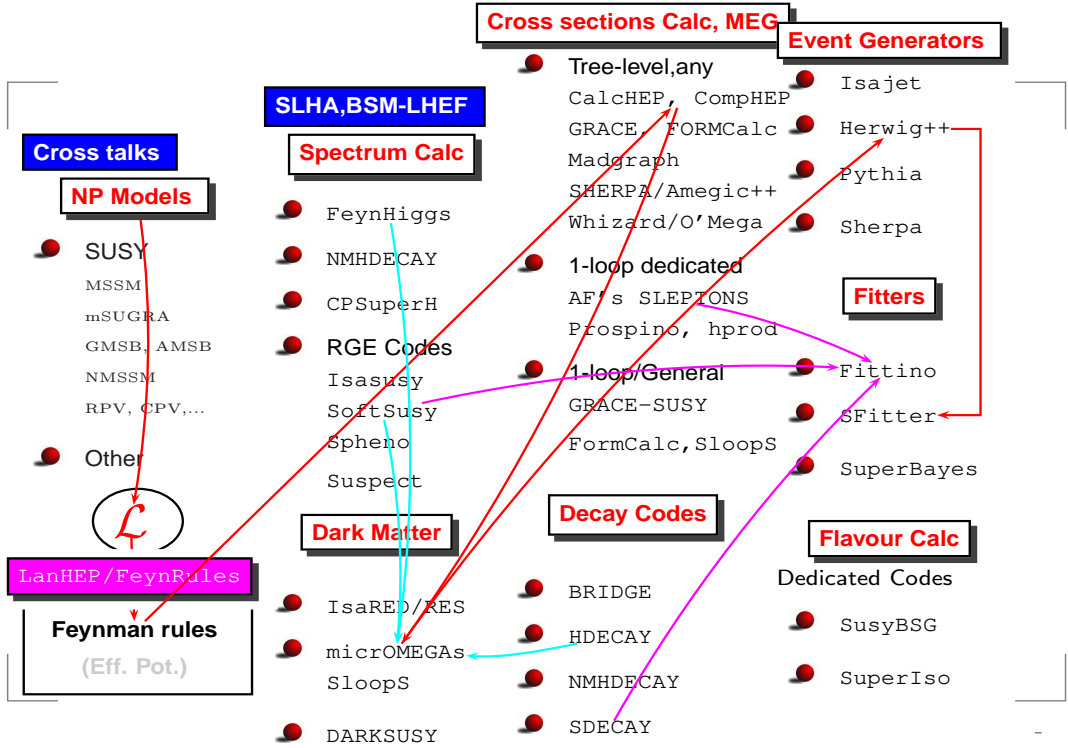}
\caption{\label{codes-bestiary}{A collection of codes for the
calculation of observables for supersymmetric particles. The event
generators and the tree-level matrix elements codes cover also the
Standard Model and in some instances other models of New Physics.
The arrows show how the output of one code is fed into  another code,
which can itself be a source for other codes.}}
\end{center}
\end{figure*}

To make contact with  LHC and LC observables one of course
needs matrix elements for cross sections and decays of the
supersymmetric particles. Many multipurpose matrix element
generators exist for supersymmetry. Among them, {\tt
Amegic}~\cite{Krauss:2001iv}, {\tt CalcHEP}~\cite{Pukhov:2004ca},
{\tt CompHEP}~\cite{Boos:2004kh}, {\tt
Grace-SUSY}~\cite{Fujimoto:2002sj}, {\tt Omega}~\cite{Kilian:2007gr}, {\tt Madgraph}~\cite{Alwall:2007st}.
Multipurpose matrix element generators can return results for any
cross section or decay of a supersymmetric particle at the tree level.
More dedicated and specialised codes in this category (cross
sections and decays) usually improve by going beyond tree-level,
among them {\tt PROSPINO}~\cite{Beenakker:1996ed} for the
production of superparticles at a hadronic collider, {\tt
HDECAY}~\cite{Djouadi:1997yw} for the decay of the Higgs and {\tt
SDECAY}~\cite{Muhlleitner:2003vg} for the decay of other
sparticles. Automatic codes for generic one-loop cross sections and decays with
supersymmetry have also been completed recently as concerns the
electroweak corrections and some classes of QCD corrections: {\tt
Grace-susy-1loop}~\cite{Fujimoto:2007bn} and {\tt
SloopS}~\cite{Baro:2007em,Baro:2008bg}. \\

For simulations at the colliders one still needs to incorporate
the matrix elements for the production (the hard
process) and the decays (of the unstable superparticle resonances)
into fully fledged Monte-Carlo generators. The latter include (i) parton shower (radiation), (ii) multiple interaction and
beam remnant in hadronic machines and (iii) hadronisation.
The main Monte-Carlo event generators are currently {\tt
Herwig++}~\cite{Bahr:2008pv}, {\tt ISAJET}~\cite{Paige:2003mg}, {\tt
Pythia}~\cite{Sjostrand:2006za} and {\tt
SHERPA}~\cite{Gleisberg:2008ta}. Fig.~\ref{mc-generators} shows the
ingredients that go into a Monte-Carlo event generator. The mass
spectrum module is what defines the model here, its content is
therefore also encoded in the dark matter codes.
One can use these generators to simulate the signatures of a
particular model at the colliders and combine these findings with
the manifestation of the same model in dark matter searches
(direct and/or indirect), the prediction of the relic density in a
particular cosmological model. One can  constrain or reconstruct
 the model even further by taking into account observables from
indirect precision measurements encoded in the flavour
calculators. Codes (``Fitters'') that perform these fits or
constraints have been written specifically with supersymmetry in
mind; we can mention {\tt Fittino}~\cite{Bechtle:2004pc}, {\tt
SFitter}~\cite{Lafaye:2004cn} and {\tt
SuperBayeS}~\cite{deAustri:2006pe}.

As we have seen, there is a very large variety of codes that cover
different aspects of the phenomenology of supersymmetry. A recent
compendium of these codes can be found in~\cite{Allanach:2008zn}.
Because of the large number of parameters in a general
supersymmetric model and because many modules are fed into other
modules,  it is best to avoid errors as much as possible when
passing parameters from one code to another. Some of these
errors can be as trivial as a problem of sign convention. The SUSY
Les Houches Accord (SLHA)~\cite{Skands:2003cj} allows an easy
parsing for the MSSM parameters. This accord has been extended
~\cite{Allanach:2008qq} to deal with more general supersymmetric
models, like the NMSSM for which a version of {\tt micrOMEGAs}
exists that uses or can be used with the {\tt NMHDECAY}
code~\cite{Ellwanger:2004xm}, or with the inclusion of CP violation
via {\tt CPSuperH}~\cite{Lee:2003nta}.

Fig.~\ref{codes-bestiary} shows how the different codes for the
calculation of supersymmetric observables both at the colliders
and in dark matter searches or the evaluation of the relic
density are interrelated. The calculation of the matrix elements
needed for these codes requires first of all reading the Feynman
rules. This in itself is a titanic endeavour  because of the
complexity of the MSSM and its extensions. For the dark matter
codes, where a very large number of processes are involved,
especially in the calculation of the relic density, practically
the whole set of rules is called for. This is even more so for
one-loop calculations. Special tools now exist to achieve this
task whereby the model file (containing the Feynman rules) is
generated automatically from just coding the Lagrangian in a
manner as close to the calculation by hand. This was first done
more than 10 years ago by {\tt LanHEP}~\cite{Semenov:1998eb} based
on $C$ for easy interface with {\tt CompHEP}.  The implementation
here is very similar to the canonical coordinate representation.
Use of multiplets and the superpotential is built-in to minimize
human error. The ghost Lagrangian is derived directly from the
BRST transformations. Very recently {\tt
FeynRules}~\cite{Christensen:2008py} based on {\tt Mathematica}
enables to perform the same task and can output to different
matrix element generators. {\tt MicrOMEGAs} rapid development was
in large part made possible because of the extensive automation
based on {\tt LanHEP} and {\tt CalcHEP/CompHEP}. {\tt LanHEP} has
also been greatly extended to automatically implement a model file
for calculations at one-loop. It is now possible  to shift fields
and parameters and thus generate counterterms most efficiently.
{\tt LanHEP} has been successfully interfaced with the highly
efficient and automatised one-loop packages based on {\tt
Feynarts/FeynCalc/LoopTools}~\cite{Hahn:1998yk,Hahn:2000kx,Hahn:2001rv,Hahn:2000jm}.
The {\tt SloopS} package is the combination of  {\tt LanHEP} and
{\tt FeynArts}, after a fully consistent and complete
renormalisation of the MSSM has been
completed~\cite{Boudjema:2005hb,Boudjema:2006kn,Baro:2007em,Baro:2008bg}.
{\tt SloopS} has been developed from the outset such that it is
applicable to both high energy collider observables and for
processes occurring at very low velocity such as is the case with
dark matter particles. This will bring the cross breeding between
dark matter and the collider predictions to  a new level of
precision, at least as far as the particle physics component is
concerned.

Combining codes to be able to conduct global analyses for
dark matter searches and the determination of its microscopic
properties are witnessing an intense activity. If future colliders
discover SUSY particles and probe their properties, one could
predict the dark matter density and would constrain cosmology with
the help of precision data provided by WMAP and PLANCK. It would
be highly exciting if the precision reconstruction of the relic
density from observables at the colliders does not match PLANCK's
determination. This would mean that the post-inflation era is most
probably not radiation dominated (see Chapter~7~\cite{Chap:Gelmini}
for a discussion of alternative cosmologies before Big Bang
nucleosynthesis and their effect on particle dark matter).
The same collider data on the
microscopic properties of DM, when put against a combination of
data from direct/indirect detection, can also give strong
constraints on the astrophysical properties of DM such as its
distribution and  clustering. These properties reveal much about galaxy
formation~\cite{Allanach:2004xn,Baltz:2006fm}. For this program to
be carried through successfully, tools developed in the
cosmology/astrophysics community and tools developed within the
particle physics collider community (and their interfaces) are
essential.

\bibliographystyle{unsrt-arxiv}
\bibliography{boudjema_edsjo_gondolo}
\end{document}